\documentclass[10pt,journal,compsoc]{IEEEtran}
\usepackage{multirow}

\ifCLASSOPTIONcompsoc
  \usepackage[nocompress]{cite}
\else
  \usepackage{cite}
\fi
\usepackage{amsmath,amssymb,amsfonts}
\usepackage{algorithmic}
\usepackage{graphicx}
\usepackage{textcomp}
\usepackage{xcolor}
\def\BibTeX{{\rm B\kern-.05em{\sc i\kern-.025em b}\kern-.08em
    T\kern-.1667em\lower.7ex\hbox{E}\kern-.125emX}}

\usepackage{amsmath}
\usepackage{cite}
\usepackage{graphicx,amssymb}
\usepackage{amsmath,amsfonts,amssymb}

\usepackage{comment}
\specialcomment{warning}{\color{red}}{\color{black}}
\specialcomment{new}{\color{blue}}{\color{black}}
\specialcomment{longcomment}{\color{gray}}{\color{black}}
\specialcomment{remove}{}{} \excludecomment{remove}

\usepackage{url}
\usepackage{graphicx}

\ifCLASSINFOpdf
\else
\fi
\usepackage{ragged2e}

\begin{document}
%
\title{Space-Time- and Frequency- Spreading for Interference Minimization in Dense IoT}
%
%
%
%

\author{Indrakshi~Dey,~\IEEEmembership{Senior Member,~IEEE,}
        and~Nicola~Marchetti,~\IEEEmembership{Senior Member,~IEEE}
\IEEEcompsocitemizethanks{\IEEEcompsocthanksitem I. Dey is with CONNECT, Department
of Electronic Engineering, Maynooth University, Ireland. 
E-mail: indrakshi.dey@mu.ie\\
N. Marchetti is with CONNECT, School of Engineering, Trinity College Dublin, Ireland. 
E-mail: nicola.marchetti@tcd.ie\\
\copyright 2022 IEEE. Personal use of this material is permitted. Permission from IEEE must be
obtained for all other uses, in any current or future media, including
reprinting/republishing this material for advertising or promotional purposes, creating new
collective works, for resale or redistribution to servers or lists, or reuse of any copyrighted
component of this work in other works.}}

%
%

\markboth{Accepted to IEEE IoT Magazine, October 2022}%
{Shell \MakeLowercase{\textit{et al.}}: Space-Time- and Frequency- Spreading for Interference Minimization in Dense IoT}
%



\IEEEtitleabstractindextext{%
\justifying
\begin{abstract}
In this article, we propose a space spreading-assisted framework that leverages either {time or frequency diversity or both to reduce interference and signal loss} owing to channel impairments and facilitate the efficient operation of large-scale dense Internet-of-Things (IoT). Our approach employs dispersion of data-streams transmitted from individual IoT devices over indexed space-time (ST), space-frequency (SF) or space-time-frequency (STF) blocks. As a result, no two devices transmit on the same block; only one is activated while the rest of the devices in the network is silent, thereby minimizing possibility of interference on the transmit side. On the receive side, multiple-antenna array ameliorates performance in presence of channel impairments while exploiting array-processing gain. As interference due to superposition of multiple data-streams is killed at its root, no extra energy is wasted in fighting interference and other impairments, thereby enabling energy-efficient transmission from multiple devices over multiple access channel (MAC). To validate the proposed concept, we simulate the performance of the framework against dense IoT networks deployed in generalized indoor and outdoor scenarios in terms of probability of signal outage. Results demonstrate that our conceptualized framework benefits from interference-free transmission as well as enhancement in overall system performance.
\end{abstract}

}

\maketitle

\IEEEdisplaynontitleabstractindextext

%
\IEEEpeerreviewmaketitle

\IEEEraisesectionheading{\section*{Introduction}\label{sec:introduction}}

%
%
%
%
\IEEEPARstart{I}{nternet}-of Things (IoT) is rapidly emerging as an advanced ecosystem of massive number of devices each with its own computational and decision-making capabilities. With proliferation through every walk of life, IoT networks promise to deliver unmatched global coverage, scalability and flexibility to handle different requirements for a comprehensive range of functionalities. This digital revolution is going to accelerate with the push to digitise industries and network operators' growing interest on expanding their businesses beyond cellular and broadband services. According to the International Data Cooperation (IDC), 41.6 billion IoT devices will be deployed worldwide by 2025 \cite{2}. Such devices will be {small size}, weight and power (SWaP) and low complexity, and include sensors, meters, wearables and trackers, such that huge volumes of them can be densely deployed within a small area in challenging radio conditions. 

\subsection*{Challenges in Dense IoT}

In a traditional network, every device can communicate over dedicated orthogonal channels and spectral resources. That is no longer a solution with hundreds of devices coexisting in a dense IoT network, as the bandwidth requirement increases exponentially with the increase in the number of devices. In  such a scenario, all devices communicate simultaneously over a shared spectrum and multi-access channel (MAC) \cite{3}. The gateway (or decision-making center) receives multiple data-streams superposed in time. 

\subsubsection*{Interference over MAC}

A plaguing challenge stemming from the above situation is the intrinsic interference resulting from the partial overlapping of multiple data-streams transmitted from multiple IoT devices at the same time and over closely-spaced frequency sub-bands. Several techniques have been proposed for interference management in IoT networks. For example, \cite{4} combines On-Off Keying (OOK) with differential coding to transmit data and variation in received power level is measured for detecting data in an interference-rich environment. Considering the downlink signal to be stronger than the reflected signal, data is detected in \cite{5} by implementing successive interference cancellation technique. Removing self-interference can sometimes degrade receive signal quality. To circumvent that problem, optimized link layer design in proposed in \cite{6}. However, none of the techniques in the literature are capable of handling interference among transmissions from heterogeneous IoT devices, like, light sensors, biosensors, actuators etc., and mitigating residual self-interference. Besides, none of these cancellation techniques consider energy-sustainability and battery-efficiency. This brings us to the second critical challenge in IoT networks; \emph{the Energy Problem}.

\subsubsection*{Energy Constraints}

{IoT devices are energy-constrained and transmitting information (data-streams) with high signal power is exacting on the battery life of the devices}. Therefore, it is not recommended for the IoT devices to transmit with a power high enough to compensate for signal losses incurred by the interference-rich environment and stringent resource constraints. Specific applications in which the devices are difficult to reach or difficult to replace or resupply with energy (e.g. sensors embedded in concrete structures \cite{7} or spread over agricultural areas or in the forest), call for a careful analysis of lifetime-extending strategies, identifying the key components of energy efficiency. Aside from designing more energy-efficient devices, or designing specific protocols to enhance energy conservation, a simple yet effective solution can be to \emph{activate a device only when and where it is necessary to transmit}.

\subsubsection*{Channel Impairments}

The wireless MAC is not only interference-rich, it also suffers from random time-varying fading and shadowing. Both the small-scale and lage-scale channel effects along with receiver noise, contribute uncertainties in the received data-streams. In addition device noise and background noise will also contribute to uncertainties in the transmitted data-streams. Moreover, specific environments like industrial ones suffer from additional impulsive noise owing to a large floor-area and presence of noisy equipments. 

Employing multiple antennas at the receive side or at the IoT gateway can improve data detection performance in presence of fading, shadowing, scattering and noise. The presence of multiple antennas allows utilization of array-processing gain to enhance fading mitigation and spectral efficiency \cite{8}. However, accommodating multiple antennas is not enough to minimize damage due to interference. Space-time coded (STC) transmission along with multiple antennas at the receiver can combat interference but at the expense of high encoding and decoding complexities \cite{9}, which, in turn, is energy- and spectrum-unsustainable. Furthermore, such techniques require detailed system knowledge on channel parameters and local observations of the IoT devices. Such observations by the devices on arrival at the receiver will be highly corrupted by the noisy environment between the devices and the receiver.

\subsection*{Contribution}

In this article, we propose a novel framework for large-scale dense IoT networks capable of addressing the interference, channel impairments and energy problems by exploiting the dimension of space while tuning either time or frequency or both. The idea is to spread transmitted data-streams over space by mapping them on indexed space-time (ST), space-frequency (SF) or space-time-frequency (STF) blocks by multiplying individual data-streams with different dispersion vectors. Consequently, only one IoT device is activated on a particular indexed block when all the other devices in the network are silent. Since no two devices are active on the same indexed slot, possibility of interference between data-streams transmitted by them is minimized. As a result, no extra energy is wasted to deal with interference, thereby materializing the possibility of low power transmission and battery-life extension at a reduced complexity. Also, worth-mentioning here is that the ST and the SF blocks will be of fixed duration and fixed bandwidth respectively. 

The receiver needs to estimate or to gain knowledge on the length of the block only once to decode or demap the transmitted signal. On the receive side, the gateway will be equipped with multiple antennas and sub-optimum multi-user detection techniques will be used. Combining STF spreading (STFS) with multi-antenna gateway (receiver) will provide considerable reduction in vulnerability to noise, fading and other channel impairments. Our proposed framework leverages the benefits of spreading, space diversity and array-processing gain to enable energy efficiency within dense IoT networks, while combating channel impairments and minimizing interference. {It is worth-noting here that by `interference', we are referring to the intrinsic interference inherent to MAC resulting from partial overlap of multiple links carrying concurrent datastreams; sometimes we generalize as \emph{inter-link interference (ILI)}.}

{The spreading we design in our research is conceptually different from traditional space-time-frequency (STF) spreading applied to multiple-access and multiplexing techniques like code division multiple access (CDMA) and orthogonal frequency division multiplexing (OFDM) respectively \cite{16}. In our proposed technique, the space dimension is created by the more than one individual IoT nodes, while the time or frequency dimension is exploited by transmitting signals from individual nodes over different time periods or frequency sub-bands. So our proposed design does not spread one received signal over different time or frequency blocks, as is done in traditional methods; we rather collect different signals from different nodes over different resource blocks.}

In our proposed method, the space dimension is created by the more than one individual IoT nodes/devices, while the time or frequency dimension is exploited by transmitting signals from individual nodes over different time periods or frequency sub-bands. So our proposed design does not spread one received signal over different time or frequency blocks, as is done in traditional method; we rather collect different signals from different nodes over different resource blocks.

In order to evaluate the advantages of our proposed framework, we employ it and analyze system performance against a scenario where devices are scattered over an annulus of area approximately equal to $3\times 10^6$ square-meters with devices deployed within distances of 100 meters of each other. {The first set of results are generated to compare outage probability performance of our proposed framework as a function of overall system SINR. It is worth-mentioning here that we refere to outage probability as the SINR point at which the received signal power at the gateway falls below the minimum detectable threshold, such that the signal power gets obscured by the interfering and noise power experienced within the dense network. We can refer to this scenario as a particular IoT node being out of the gateway range. The second set of results demonstrate how interfering power varies with signal power transmitted by the IoT nodes. Our spreading framework offers a reduction of interference by around 5-8dB compared to the traditional massive IoT networks. The third and forth set of results demonstrate that interfering power increase rate slows down by employing the SRFS framework and that an increase in the number of antennas at the gateway improves the overall system SINR, respectively.}

\section*{STF Spreading Framework}

\begin{figure*}[t]
\begin{center}
 \includegraphics[width=0.85\linewidth]{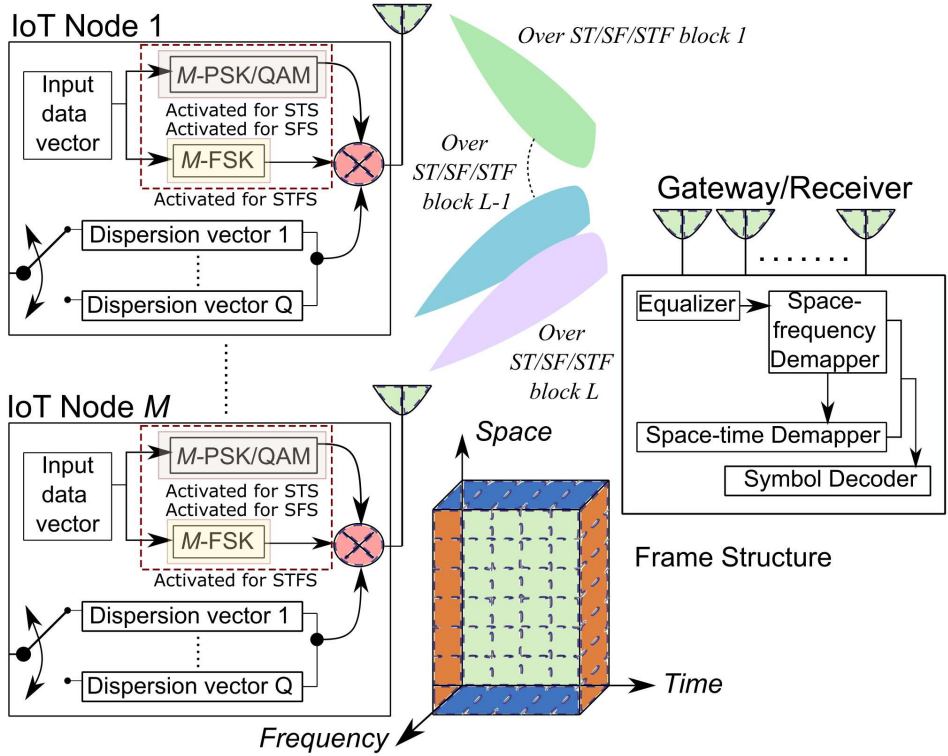}
\end{center}
\vspace*{-122mm}
\caption{The envisioned framework of a dense IoT network aided with STF spreading at the transmit side and multi-antenna gateway on the receive side for interference minimization within the network.}
\label{FIG1}
\vspace*{-5mm}
\end{figure*}

The envisioned framework is presented in Fig.~\ref{FIG1}, where the data-streams from each of the IoT devices are mapped onto fixed-length dispersion vectors before being transmitted over an indexed ST, SF or STF block. The concept of attaching indices to the ST/SF/STF mapping blocks enables the possibility of activating a subset of all the available set of different parameters, for example, only a subset of all the dispersion vectors generated for actual mapping and transmission. This indexing also allows sending additional information beyond what is allowed by traditional coding and modulation techniques, like, which device is to be activated, or what device activation pattern is to be incorporated.

The data-streams are transmitted simultaneously over an orthogonal and a coherent MAC. This means that the transmitted data-streams are travelling over channels that are coherent in frequency but may not be coherent in phase. This implies that we need to either implement distributed phase synchronization or the devices need the knowledge of the phase of the complex channel coefficients, so that they can cancel them before transmission. However, phase shift keying (PSK) can be implemented as it does not need extra phase synchronization at low data rates ($<1$ Gbps), the range in which most IoT devices operate at.

The framework allows the network to exploit time, frequency and space diversity by tuning five different parameters, i) Length of the dispersion vectors ($T$), ii) Number of dispersion vectors from which a suitable vector can be chosen randomly or depending on a particular design criteria ($Q$), iii) Number of ST or SF or STF blocks that the total allowable spectrum and time period of transmission can be divided into ($L$), iv) Number of devices that are allowed to transmit simultaneously ($M$) and v) Number of the antennas active on the receive side ($N$). Depending on the kind of application, the network topology (star, mesh, wheel etc.) and the resources available, all or some of the parameters can be tuned to exploit different degrees of freedom in the network. This framework is inspired by our seminal work in \cite{10} where we applied ST spreading (STS) to massive wireless sensor networks (WSNs) to improve performance of distributed detection and decision-fusion. Such a framework also includes concepts of spatial modulation (SM) \cite{11}, space shift keying (SSK) \cite{12} and space time frequency shift keying (STFSK) \cite{13} as special cases. 

Our proposed framework exploits space diversity in conjunction with time and frequency diversity. Space diversity commonly refers to antenna diversity, where more than one antenna is used to transmit and receive and the antennas are separated in space. The signals either transmitted or received on different antennas are uncorrelated in space. In our framework, IoT devices are involved as transmitters, where in most cases, the devices are low SWaP equipped with a single antenna. As the devices are separated in space, transmit space-diversity is automatically leveraged. 

Moreover, as the devices never fire in the same time slot or same frequency slot, the transmitted signals are separated by time and frequency, thus offering time and frequency diversity. In dense networks, however, it is very important to choose the correct time separation and frequency separation. For example, if we choose ST blocks of duration of $10~\mu$s and there are 100 devices active within the network, we will experience a transmission delay in the range of 1 - 10 ms. This amount of delay is tolerable to certain massive IoT network applications. However, such a delay can be challenging to compensate for critical IoT-like networks, especially in healthcare applications. In that case, frequency diversity needs to exploited. We also need to be careful about the frequency separation. For example, if we choose separation between SF blocks to be equal to a minimum of 10Hz, with 100 devices in the network, we need a minimum of 1 KHz of spectrum dedicated to guard bands only. In that case, the total available bandwidth for the network will guide the choice of the size of the SF blocks and the separation between them.

\subsection*{Transmitter Design}

Each data-stream is modulated using either $M$-phase shift keying ($M$-PSK)/ $M$-Quadrature Amplitude Modulation ($M$-QAM) or $M$-frequency shift keying ($M$-FSK) before being mapped on the ST blocks or the SF blocks respectively. For the STFS arrangement, each data-stream is divided into two parallel bit-streams, the first one is modulated using $M$-PSK/$M$-QAM and the second one with $M$-FSK. The two modulated streams are combined before being mapped on the STF blocks.

\subsubsection*{{Tuning with time}} 
If we want to exploit only the time diversity, the transmission frame is divided into several ST blocks. Each IoT device is allowed to transmit on a dedicated ST block over which other devices in the network will remain silent. All the devices are allowed to transmit simultaneously over the entire channel bandwidth. Possible scenarios where this arrangement would be apt are the bandwidth-hungry applications, like autonomous driving networks, or fully immersive Augmented Reality (AR)/ Virtual Reality (VR), where the devices need a larger bandwidth to transmit. However, the devices do not transmit at the same time and therefore, they do not interfere with each other, even if they transmit over the same bandwidth.

\subsubsection*{{Tuning with frequency}} 
The transmission frame in this case can be divided into individual SF blocks where each IoT device can transmit over individual frequency sub-bands. All the devices can transmit simultaneously at the same time on different frequency sub-bands. In each SF block, only one device is allowed to be active while the other ones are silent. This technique will be suitable for time-sensitive IoT networks capable of delivering e.g. remote healthcare applications or intelligent transportation systems.

\subsubsection*{{Tuning both time and frequency}}
Data-streams from individual IoT devices are transmitted over individual STF blocks; a technique coined as STFS. The transmission frame is divided into indexed ST and SF blocks of fixed duration and bandwidth respectively. This arrangement can be applied to dense IoT networks delivering time-sensitive high-throughput applications, like semi-autonomous or remote-controlled robots in industrial settings that are supposed to handle large volumes of data e.g. real-time video, or high resolution sensor observations. 

\subsection*{Dispersion Vectors}

A crucial component of this framework is the design of the dispersion vectors, where each of the vectors satisfies the power constraint. The power constraint refers to the condition where the trace of the product of each vector with its complex conjugate transpose is equal to the length of the vector. Each of these vectors includes only one single non-zero component. A set of $Q$ dispersion vectors can be generated randomly using Gaussian distribution. A possible good set of dispersion vectors result in low correlation between the obtained transmitted signals. The vector generation can be repeated to increase the set of choices from which the optimum array of vectors can be selected. The set of vectors can be chosen based on the overall i) minimization of the detection error probability or ii) maximization of link capacity of the network. Once the design criterion is selected, an optimum set of vectors are generated through exhaustive search. This set of vectors will be capable of satisfying the criterion and the power constraint depending on the parametric values of $M$, $N$, $L$ and $T$. The number of times the vector generation is repeated, depends on the design parameters and the operational signal-to-noise-plus-interference ratio (SINR) point at which the link capacity or the error performance is computed. 

Another way of getting optimum dispersion vector set is to select the first $T$ rows or the first $M$ columns of a set of unitary matrices for the case of $(M \geq T)$ and $(M < T)$ respectively. In order to maintain the power constraint, a constant of $\sqrt{T/M}$ should be multiplied by all the dispersion vectors. The optimum set of vectors can be obtained by maximizing the minimum distance between consecutive spread and modulated transmit signals. {Therefore, the choice of the dispersion vectors used for spreading depends on the estimated received information vector, which in turn, varies with the modulation scheme used for mapping the transmit datastreams.}

\subsection*{Receiver Design}

When the transmit data-streams are ST spread, an ST demapper is implemented on the receive side to identify the activated time-slots, followed by a symbol decoder that can estimate the dispersion vector and received data from the demapped signal. A frequency demapper followed by a symbol decoder will be implemented on the receive side for the SF spread transmission arrangement. If both ST and SF spreading are employed, the gateway (or receiver) will apply a SF demapper first, followed by a ST demapper and a symbol decoder. 

If the frequency dimension is exploited, it is recommended to employ any form of channel equalizer on the receiving side before SF demapping, in order to eliminate an inherent delay-distortion between the frequency components. Symbol decoder can take any form depending on the application, resource availability and complexity tolerance of the receiver end, including maximum likelihood (ML), minimum mean-squared error (MMSE), zero-forcing (ZF), matched filter or other Bayesian techniques. 

The receiver should also have access to estimated channel state information (CSI) to decode the data streams. Therefore, a part of the channel coherence time needs to be used for training. During the training phase, the devices will simultaneously transmit mutually orthogonal pilot sequences over the allocated ST, SF or the STF blocks for individual devices. Once all the pilot sequences are collected at the receiver, any channel estimation algorithm like MMSE or least square (LS) can be used to estimate the channel states.

{Allocating mutually orthogonal pilot sequences for estimating individual CSIs is challenging within massive and dense IoT networks owing to the following reasons; i) the number of generable orthogonal pilot sequences is limited, ii) reusing pilot sequences is not feasible owing to pilot contamination when multiple nodes are transmitting at the same time or on the closely-spaced frequency sub-bands and iii) the number of resource blocks to be allocated increases proportionally with the number of active nodes. ST/SF/STF spreading offers an elegant solution to the above challenges. Since no two devices transmit on the same ST/SF/STF block, it will be possible to reuse orthogonal pilot sequences without the possibility of pilot contamination and the need to allocate new resource blocks to cater to each devices separately.}


\section*{Numerical Performance and Analysis}

In this section, we simulate the performance of the prototype of the illustrated framework, where we choose channel parameters to represent generalized indoor and outdoor scenarios. Here, we will briefly describe the couple of scenarios considered and illustrate results for both indoor and outdoor settings. 

\subsection*{Scenarios and Metrics}

We consider two different scenarios - a) indoor, like a smart home, open office or smart industry and b) outdoor, like an autonomous vehicular network communicating over an intelligent transportation system. In both cases, the IoT devices are randomly deployed with uniform distribution over a circular annulus with the gateway (or receiver) at the center and with a maximum and minimum radii of 1 Km and 100 m respectively. The MAC is assumed to be dispersive with independently and identically distributed Rayleigh fading and log-normal shadowing envelope, while the pathloss is experienced according to the 3GPP Urban Microcell model \cite{14} for outdoor links and 3GPP Indoor Hotspot model \cite{15} for indoor links. 

STFS offers the flexibility of playing with five different parameters which need to be appropriately selected and configured based on the requirement and scenario. For a narrowband network, STS will be a better choice while for a wideband application, SFS and STFS are better options. On the other hand, the length of the dispersion vector $T$ will have a profound effect on the link capacity. Link capacity will decrease linearly with the increase in $T$. But at the same time, we cannot choose $T$ to be less than $M$ i.e. $T \geq M$, the number of devices active within the network. Similarly, the number of ST/SF/STF blocks, $L$ cannot be less than $M$ i.e. $L \geq M$. However, higher $L$ means higher computational complexity. So it is recommended that $L$ should be equal to $M$. The number of available dispersion vectors $Q$ that we can choose from also has to be at least equal to $M$, i.e. $Q \geq M$. Higher $Q$ provides higher throughput and diversity gain but at the cost of higher complexity. Generally, as system designers we may not have control on $M$ and $N$ depending on the network at hand. For example, we cannot change $M$ or $N$ within existing network architecture. If we are designing and deploying a completely new network, we can increase $M$ and $N$ to enhance diversity and multiplexing gain but at the cost of increased complexity at the receive side. 

For simulating the performance, a square-law detector is used to detect $M$-FSK symbols when SFS and STFS are employed. For detecting the $M$-FSK/$M$-QAM symbols generated as part of STS and STFS, performance of two symbol decoders are compared, ML and ZF. Small-scale statistics of the channel is modelled as Rayleigh frequency-selective distributed with 4-taps and a normalized Doppler frequency of 0.01 Hz. For the SFS and STFS set, we also employ ZF equalizer on the receive side before the demapper and symbol decoder.

\subsection*{Results and Analysis}

\begin{figure}[t]
\begin{center}
 \includegraphics[width=1.75\linewidth]{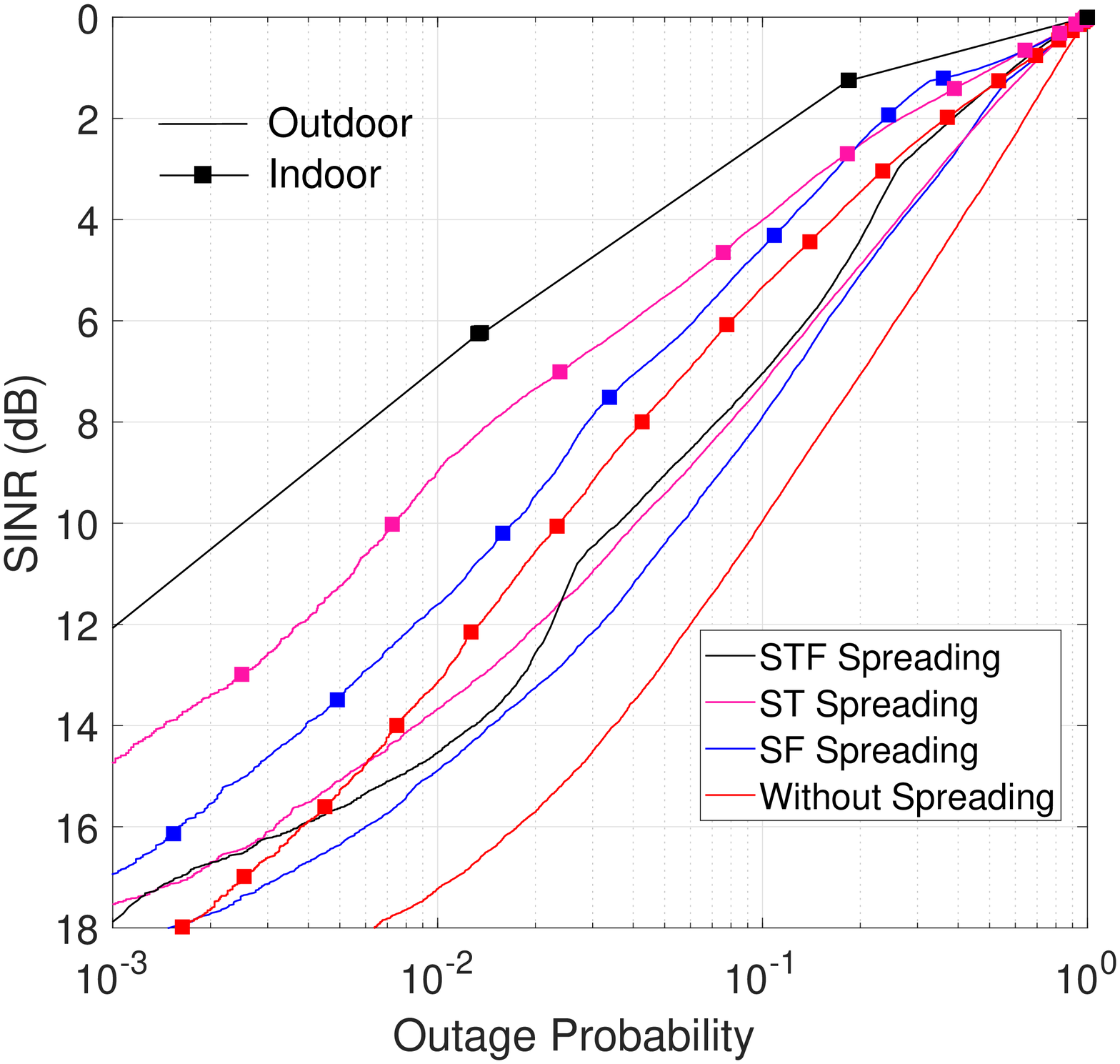}
\end{center}
\vspace*{-7mm}
\caption{Comparative outage probability performance of a dense IoT network with $M = 8$ equipped with single antennas and $N = 64$, $L = 8$, $T = 8$, $Q = 8$ as a function of SINR (dB) - all three techniques ST, SF and STF spreading are compared with the case without spreading. Both the devices  and the gateway/receiver are deployed in an outdoor or an indoor environment.}
\label{FIG3}
\vspace*{-5mm}
\end{figure}

\begin{figure}[t]
\begin{center}
 \includegraphics[width=1.8\linewidth]{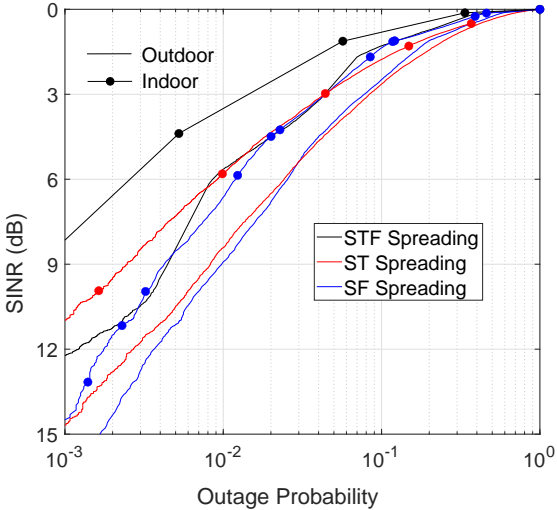}
\end{center}
\vspace*{-10mm}
\caption{Comparative outage probability performance of a dense IoT network with $M = 16$, $N = 100$, $L = 20$, $T = 20$, $Q = 20$ as a function of SINR (dB) - all three spreading techniques are compared. Both the devices and the gateway/receiver are deployed in an outdoor or an indoor environment.}
\label{FIG4}
\vspace*{-6mm}
\end{figure}

{We have generated three sets of results comparing the performances of STS, SFS and STFS techniques with different sets of $(M, N, L, T, Q)$ for two different scenarios; outdoor with 3GPP Urban Microcell pathloss model \cite{14} and indoor with 3GPP Indoor Hotspot pathloss model \cite{15}. The first result (Fig.~\ref{FIG3}) demonstrates outage probability performance as a function of signal-to-interference-plus noise ratio (SINR) for the case where $M = L = T = Q = 8$. The second one (Fig.~\ref{FIG4}) presents a comparative probability of data outage again as a function of SINR for the case where $M = 16$ and $L = T = Q = 20 > M$. In both of the above-mentioned cases, the number of antennas on the receive side $N = 64$ and $100$ are much larger than the number of IoT devices present in the network, i.e. $N \gg M$ in order to leverage the array processing gain. In both cases, we have used ML symbol detection on the receive side. Both the scenarios experience moderate shadowing with mean 4dB and variance 2dB.}

Plotting outage probability is an important metric for providing us a vivid picture of how individual links within a dense IoT network platform perform. In our case, outage probabilities are the points at which received signal power over individual links between the IoT devices and the receiver falls below the threshold of 0 dB. This means that the devices will be out of range of the gateway/receiver, owing to the interfering power from other devices overshadowing actual transmit signal power from each active device. In both cases, STF spreading offers significant performance gain in indoor environments. In outdoor scenario, STF and ST spreading offer almost equivalent performance for $M = L = T = Q$. However, STF offers gain in performance over ST spreading if $L = T = Q > M$ but at the cost of higher computational complexity. 

\begin{figure}[t]
\begin{center}
 \includegraphics[width=0.99\linewidth]{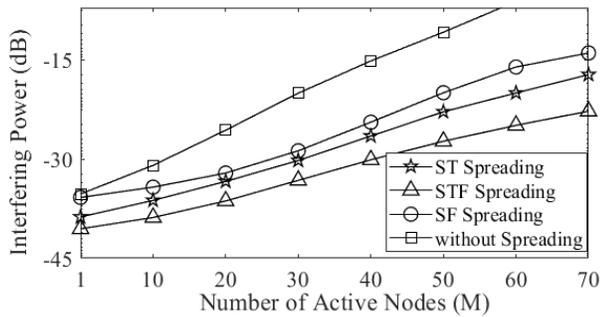}
\end{center}
\vspace*{-8mm}
\caption{{Demonstration of how interfering power changes with number of active nodes in the network ($M$) with $N = 64$, $L = T = Q = 8$ in an indoor scenario.}}
\label{FIG5}
\vspace*{-3mm}
\end{figure}

{Fig.~\ref{FIG5} demonstrates how interfering power varies when the number of nodes active within the network increases. Assuming all the nodes transmit with equal power in an indoor scenario with $L = Q = T =8$, $N = 64$, interfering power experienced within the network increases with the number of nodes transmitting. It is worth-noting here that the interfering power is calculated by adding up the average and the mean-squared delay spreads experienced over the channel taps. However, the rate of increase in interfering power slows down considerably from no spreading scenario to ST/SF spreading-aided cases and then to STFS-aided transmission. For example, if $M = 40$, total interference seen is -15dB without spreading and -32dB with STFS implemented, thereby lowering the probability of interference by around 50\%.}

\begin{figure}[t]
\begin{center}
 \includegraphics[width=0.99\linewidth]{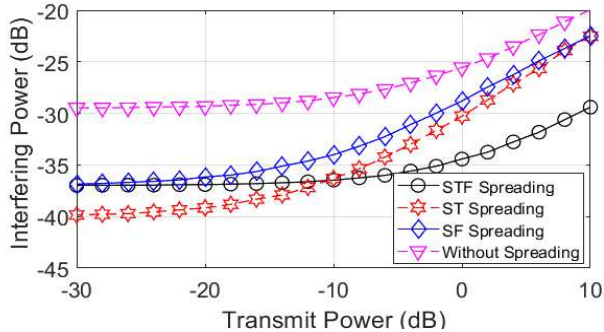}
\end{center}
\vspace*{-5mm}
\caption{Representation of how interfering power changes with the transmit power within a dense IoT network with $M = 8$, $N = 64$, $L = 8$, $T = 8$, $Q = 8$ where the devices are deployed within a smart home like indoor environment.}
\label{FIG6}
\vspace*{-5mm}
\end{figure}

The set of curves plotted in (Fig.~\ref{FIG6}) exhibits how interfering power varies with signal power with which the devices within a dense IoT network transmit their data-streams. This final set of results is generated assuming that all the devices transmit with equal power, and is used to visualize how interference varies with transmit signal power when the spreading framework is employed in an indoor scenario with $M = L = Q = T = 8$ and $N = 64$. Interfering power increases with the increase in transmit power in all possible scenarios. When ST/SF/STF spreading is employed, the amount of interference resulting within the network decreases by around 5 - 8 dB. ST offers best interference minimization at lower transmit powers, but as the transmit power range increases, STF emerges to be the winner. As the end-to-end signal power decays, the reflected signal power contributed by the multipath components arriving at the receiver manifests and results in further signal loss. The number of multipath components arriving at the receiver are more in case of ST spreading than SF spreading. Interfering power therefore increases faster with ST spreading than with SF spreading. 

\begin{figure}[t]
\begin{center}
 \includegraphics[width=0.99\linewidth]{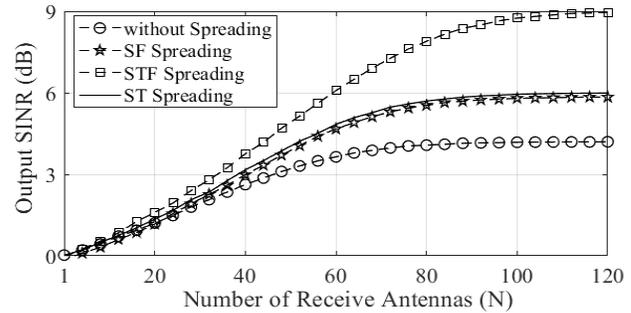}
\end{center}
\vspace*{-5mm}
\caption{{Comparative output SINR of an IoT network with the increase in the number of receive antennas at the gateway ($N$) with $M = L = T = Q = 8$ where the devices are deployed in an indoor environment.}}
\label{FIG7}
\vspace*{-5mm}
\end{figure}

{The ratio of the receive energy of the serving signal to the total interfering power plus simulated noise experienced at the output of the gateway is plotted as the output SINR in Fig.~\ref{FIG7}. Essentially, the variation of the overall SINR at the output of the gateway is plotted as a function of the number of the receive antennas ($N$) at the gateway for an indoor scenario with all nodes transmitting with equal transmit power. Increase in $N$ does provide an advantage with improving the overall SINR. However, that improvement in SINR saturates with $N > 60$. The point of saturation increases from SF to ST to STF which means that if STF is employed, the performance gain can be maintained with increased $N$ to a considerably large-sized antenna-array.}

\section*{Conclusion and Discussion}

In order to strike a flexible balance between interference minimization and energy efficiency in massive IoT networks, we have conceptualized the framework of STF spreading of the transmitted data-streams before transmission over a coherent MAC. The resultant framework will not only benefit from improvement in opportunistic throughput but also from interference-free transmission within a dense IoT network. We tuned different parameters like number of devices active in the network, number of receive antennas, length and number of ST/SF/STF blocks, as well as the number of dispersion vectors available for spreading. With higher values of the parameters, enhancement in throughput is achievable, however at the cost of higher computational complexity. Therefore, the choice of the parameters and achievable throughput needs to be jointly optimized to attend a perfect balance, such that the operation is not complex enough to be taxing for the IoT devices in terms of energy consumption.

Our framework has been designed and evaluated primarily keeping stationary devices or mobility-constrained devices in mind. However, this framework should also be applicable to mobile or highly dynamic entities. In such a dynamic scenario, a major concern will be the short channel coherence time over which the devices have to transmit information as well as training symbols. A possible way forward can be to introduce the concept of differential MAC and differential spreading of the transmitted data-streams.

Another important aspect is the number of antennas at each of the IoT devices active within the network. As of now, we are considering devices equipped with a single antenna. Hence we accounted for the number of active devices = number of transmit antennas = $M$. However, in a heterogeneous IoT network, it may happen that some devices are single-antenna while others are equipped with two or more antennas. In that case, the framework has to be extended to take care of dissimilar number of transmit antennas at the devices. A possible solution can be to include another parameter like antenna pattern diversity. Instead of mapping indices of the devices to the ST/SF/STF blocks, multiple antenna patterns can be exploited. Several multi-antenna patterns can be generated, out of which one or a subset can be activated depending on the scenario. Information will be conveyed over these patterns and then mapped onto the spreading blocks. Another possible way is to implement multi-slot transmission, where multiple data-streams generated by multiple antennas at a single IoT device can be mapped over individual ST/SF/STF blocks.
\vspace*{-5mm}

\end{document}